\documentclass[3p,times,procedia,sort&compress]{elsarticle}
\usepackage{nupha_ecrc}

\volume{00}

\firstpage{1}

\journalname{Nuclear Physics A}

\runauth{M. Martinez, M. McNelis, and U. Heinz}

\jid{nupha}

\jnltitlelogo{Nuclear Physics A}

\usepackage{amssymb}
\usepackage{graphicx}
\usepackage{bm}
\usepackage{amsmath,latexsym}
\usepackage[usenames]{color}
\usepackage{subfigure}
\usepackage{subfigure}
\usepackage{slashed}
\usepackage{multirow,array}
\usepackage{mathtools}
\usepackage{mathrsfs}
\usepackage[colorlinks=false,linktocpage=true]{hyperref}
\usepackage{hyperref}
\usepackage[utf8]{inputenc}
\usepackage{lipsum}
\usepackage{color}
\definecolor{darkblue}{RGB}{0,0,196}
\definecolor{darkred}{RGB}{196,0,0}

\usepackage[figuresright]{rotating}
\newcommand{\be}{\begin{equation}}
\newcommand{\ee}{\end{equation}}
\newcommand{\bea}{\begin{eqnarray}}
\newcommand{\eea}{\end{eqnarray}}
\newcommand{\beal}{\begin{align}}
\newcommand{\enal}{\end{align}}
\newcommand{\bs}{\begin{subequations}}
\newcommand{\es}{\end{subequations}}
\newcommand{\besp}{\begin{split}}
\newcommand{\eesp}{\end{split}}
\newcommand{\x}{\hat{x}^\mu}

\newcommand{\tem}{\hat{T}}

\newcommand{\ene}{\hat{\epsilon}}

\begin{document}

\begin{frontmatter}

\title{Viscous anisotropic hydrodynamics for the Gubser flow}

\author[ncsu,osu]{M. Martinez}
\author[osu]{M. McNelis}
\author[osu]{U. Heinz}

\address[ncsu]{Department of Physics, North Carolina State University, Raleigh, NC 27695}
\address[osu]{Department of Physics, The Ohio State University, Columbus, OH 43210, USA}

\begin{abstract}
In this work we describe the dynamics of a highly anisotropic system undergoing boost-invariant longitudinal and azimuthally symmetric radial expansion (Gubser flow) for arbitrary shear viscosity to entropy density ratio. We derive the equations of motion of dissipative anisotropic hydrodynamics by applying to this situation the moments method recently derived by Moln\'ar et al. (MNR)~\cite{Molnar:2016vvu,Molnar:2016gwq}, based on an expansion around an arbitrary anisotropic one-particle distribution function. One requires an additional evolution equation in order to close the conservation laws. This is achieved by selecting the relaxation equation for the longitudinal pressure with a suitable Landau matching condition. As a result one obtains two coupled differential equations for the energy density and the longitudinal pressure which respect the $SO(3)_q\otimes SO(1,1)\otimes Z_2$ symmetry of the Gubser flow in the deSitter space. These equations are solved numerically and compared with the predictions of the recently found exact solution of the relaxation-time-approximation Boltzmann equation subject to the same flow. We also compare our numerical results with other fluid dynamical models. We observe that the MNR description of anisotropic fluid dynamics reproduces the space-time evolution of the system than all other currently known hydrodynamical approaches.
\end{abstract}


\end{frontmatter}

\section{Introduction}
\label{sec:introduction}
Relativistic hydrodynamics has been widely used to describe different experimental observables measured in high energy nuclear collisions. Its success has sparked the interest of the scientific community to understand the hydrodynamical behaviour of strongly coupled systems. {\it A priori} hydrodynamical models are not expected to provide a good description when the mean free path becomes comparable to the macroscopic scales of the system. This situation occurs especially at the early stages of the expansion of the fireball  when the rapid expansion of the system competes with the microscopic relaxation processes of the quark-gluon plasma. When the rate of expansion is anisotropic to continue driving the system further away from local thermal equilibrium, then this manifest macroscopically through large pressure anisotropies. Large pressure anisotropies persists during the entire expansion of the system up to scales of the order of the hydrodynamic relaxation rates. The main aim of anisotropic hydrodynamics~\cite{Martinez:2010sc,Martinez:2010sd,Florkowski:2010cf,Ryblewski:2010bs,Bazow:2013ifa,Molnar:2016vvu} is to address these type of problems and to provide a consistent framework where pressure anisotropies are taken into account in a non-perturbatively controlled way. 

Within the kinetic theory approach it is possible to derive the equations of motion of the macroscopic hydrodynamic variables (e.g. energy density) by considering the evolution equation of the slowest hydrodynamical moments of the distribution function. The standard second order viscous hydrodynamical theories are obtained by expanding around a local equilibrium distribution function which is isotropic in momentum-space. Deviations from equilibrium generate dissipative corrections which capture non-zero mean free path corrections and can be parametrized through to transport coefficients such as the heat conductivity, shear and bulk viscosities. Different methods to derive second order theories of hydrodynamics have been considered in the literature. Those methods do not necessarily all lead to the same values of the transport coefficients~\cite{Denicol:2014loa}. However, all those methods assume that the deviations from equilibrium are small. Anisotropically expanding plasmas develop large pressure anisotropies which survive during extended periods of time and thus, an expansion around the equilibrium is bound to break down and its validity must be questioned. Furthermore, the applicability of standard derivations of hydrodynamics is questionable when large pressures anisotropies are present.  

Anisotropic hydrodynamics considers an expansion around a distribution function which is anisotropic in momentum-space~\footnote{For a system where the chemical potential vanishes exactly, this distribution function is assumed to have the Romatschke-Strickland form~\cite{Romatschke:2003ms} which in the local rest frame looks like  $f (x^\mu,p_i)=f\left(\sqrt{p_T^2+(1+\xi)p_z^2},\Lambda\right)$ where $\Lambda$ is a particular momentum scale which is identified with the temperature by matching the energy density~\cite{Martinez:2010sc}.}. The amount of anisotropy along some particular direction is measured by a deformation parameter $\xi$. This parameter characterizes the leading order term of the anisotropic distribution function. In principle, it is hoped that the space-time evolution of the deformation parameter encodes the information of the macroscopic dynamical behaviour of the pressure anisotropies. Nevertheless, there is no first principles calculation which determines uniquely the evolution equation for this deformation parameter. Different prescriptions to determine the evolution equation of the deformation parameter have been proposed but none of those have achieved conceptual clarity.

Recently, Moln\'ar et. al.~\cite{Molnar:2016vvu,Molnar:2016gwq} derived anisotropic hydrodynamical equations of motion by considering the so-called $\mathcal{P}_L$ matching. Within this approach the anisotropy parameter $\xi$ is treated as a Lagrange multiplier which is determined by matching it to the longitudinal pressure $\mathcal{P}_L$. $\mathcal{P}_L$ corresponds to a certain moment of the leading order anisotropic distribution function. Its evolution equation is determined from the Boltzmann equation. As a result, this scheme matches exactly the total pressure anisotropy without requiring any input of the microscopic deformation parameter $\xi$. The numerical solutions of anisotropic hydrodynamics with the $\mathcal{P}_L$ matching scheme showed to reproduce to high numerical accuracy the predictions of the exact Boltzmann equation within the relaxation time approximation (RTA) for the Bjorken flow~\cite{Molnar:2016gwq}. 

In these proceedings we discuss some of the results presented in our recent publication~\cite{Martinez:2017ibh}. We extend and confirm the results of Moln\'ar et. al.~\cite{Molnar:2016gwq} for a rapidly expanding conformal fluid undergoing  Gubser flow~\cite{Gubser:2010ze,Gubser:2010ui}. This flow describes a conformally invariant system that expands azimuthally symmetrically in the transverse plane in addition to boost-invariant longitudinal expansion. Its evolution is more easily described in the three dimensional de Sitter space times a line $dS_3\otimes R$ where the underlying $SO(3)_q\otimes SO(1,1)\otimes Z_2$ conformal symmetry of the Gubser flow is explicitly manifest~\cite{Gubser:2010ze,Gubser:2010ui}. An exact solution to the RTA Boltzmann equation for the Gubser flow has been recently found~\cite{Denicol:2014xca,Denicol:2014tha}. This solution allows us to compute exactly the evolution of the slowest hydrodynamic moments, the energy density (and thus, the temperature) and the effective shear stress. Therefore, by comparing the results obtained from the exact solution one can test the validity and accuracy of different hydrodynamical schemes. 

\begin{figure}[h]
\centerline{
\includegraphics[width=1\linewidth]{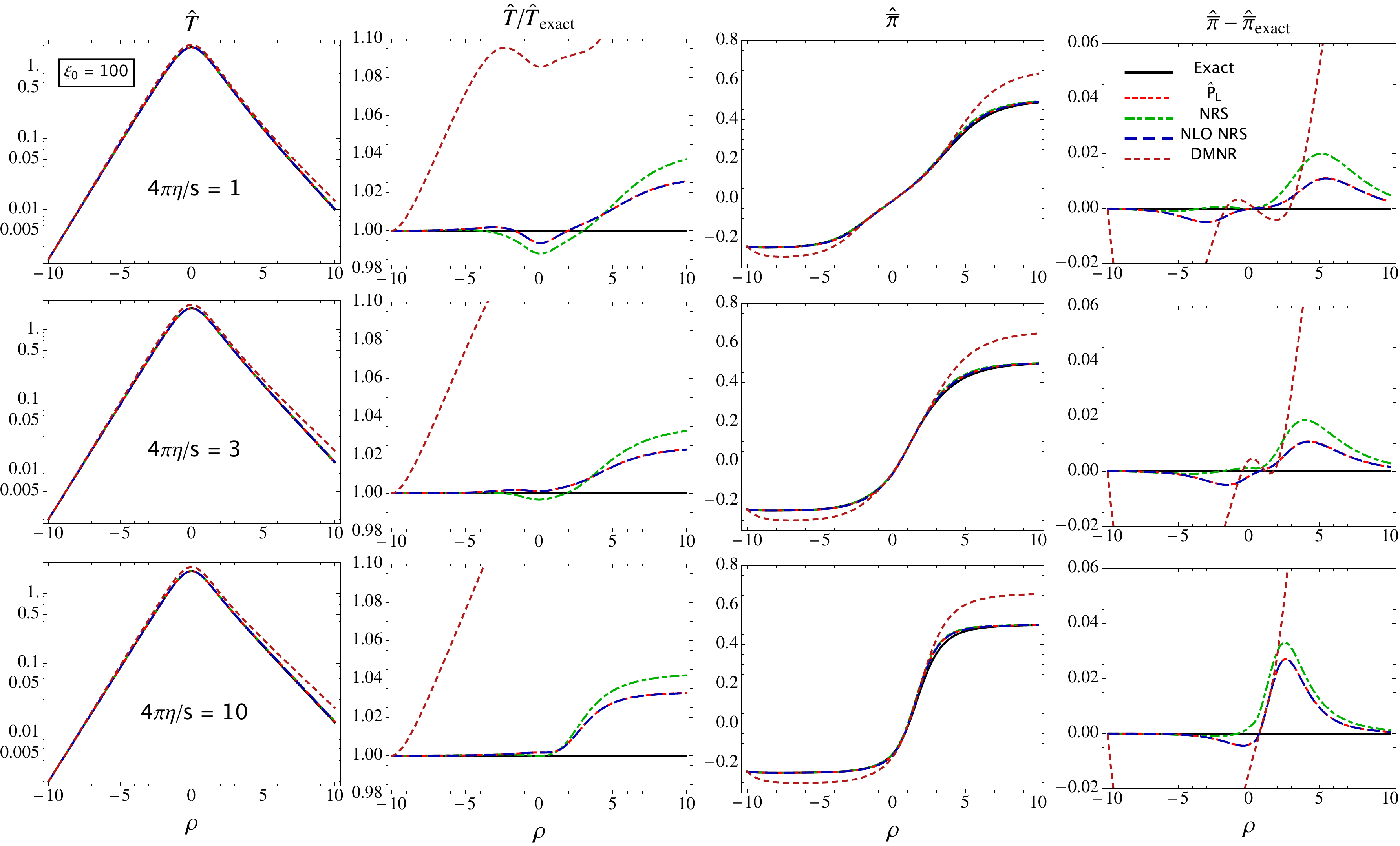}
}
\caption{
de Sitter time evolution of the temperature $\hat{T}$ and the normalized shear stress $\hat{\bar{\pi}}$ for the exact solution of the RTA Boltzmann equation~\cite{Denicol:2014xca,Denicol:2014tha} (black solid lines) and four different hydrodynamic approximations: second-order viscous hydrodynamics (DNMR)~\cite{Denicol:2012cn} (short-dashed magenta lines), anisotropic hydrodynamics with $\hat{\mathcal{P}}_L$-matching~\cite{Molnar:2016vvu,Molnar:2016gwq,Martinez:2017ibh} (dotted red lines), NRS scheme~\cite{Nopoush:2014qba} (dash-dotted green lines) and NLO NRS description~\cite{Martinez:2017ibh} (long-dashed blue lines). Figure taken from Ref.~\cite{Martinez:2017ibh}.
}
\label{fig1}
\end{figure}

\section{Results}
In our work~\cite{Martinez:2017ibh} we compare the predictions of the exact solution to the RTA Boltzmann equation for the Gubser flow~\cite{Denicol:2014xca,Denicol:2014tha} with four different hydrodynamical models: anisotropic hydrodynamics with $\hat{\mathcal{P}}_L$ matching~\cite{Molnar:2016vvu,Molnar:2016gwq,Martinez:2017ibh}, leading order anisotropic hydrodynamics in the NRS scheme~\cite{Nopoush:2014qba}, next-to-leading order anisotropic hydrodynamics in the NRS scheme~\cite{Martinez:2017ibh} amended by residual viscous corrections (NLO NRS)~\cite{Bazow:2013ifa,Martinez:2017ibh} and standard second-order viscous hydrodynamics (DNMR)~\cite{Denicol:2012cn}. We refer the interested reader to Ref.~\cite{Martinez:2017ibh} for the derivation and discussion of the corresponding equations of motion.

In Fig.~\ref{fig1} we present the numerical solutions for the de Sitter evolution of the temperature $\tem$ and the normalized shear stress $\hat{\bar\pi}=3\hat\pi/(4\ene)$ where $\ene$ is the energy density. In this figure we choose an initial anisotropic distribution function with $\xi_0=100$, initial de Sitter time $\rho_0=-10$ and initial temperature $\tem_0=0.002$. The top, middle and bottom rows of panels correspond to specific shear viscosity $4\pi\eta/s=1$, 3, and 10, respectively. The four columns of plots show, from left to right, the $\rho$ evolution of the temperature $\hat{T}$, of the ratio $\hat{T}/\hat{T}_\mathrm{exact}$, of the normalized shear stress $\hat{\bar\pi}$, and of the difference $\hat{\bar\pi}{-}\hat{\bar\pi}_\mathrm{exact}$. Additional set of initial conditions were explored in Ref.~\cite{Martinez:2017ibh}. 
 
In all the studied cases we observe that all anisotropic hydrodynamical prescriptions match the exact results better than does viscous hydrodynamics.  We observe that in the range of de Sitter time-like variable $\rho$ shown here, anisotropic hydrodynamics with $\hat{\mathcal{P}}_L$ matching scheme provides the best approximation to the exact solution. Especially at large $\rho$ values where the Knudsen number grows exponentially~\cite{Denicol:2014tha}, anisotropic hydrodynamics with $\hat{\mathcal{P}}_L$ matching deviates from the exact solution by more than a few percent while  the leading order NRS prescription~\cite{Nopoush:2014qba} does not perform as well in the asymptotic $\rho >> 0$ region~\cite{Denicol:2014tha}. Second-order viscous hydrodynamics~\cite{Denicol:2012cn} exhibit large deviations from the exact result over the $\rho$ interval studied here. These conclusions continue to hold for different initial conditions and values of the shear viscosity over entropy ratio $\eta/s$~\cite{Martinez:2017ibh}.

We also observe that the mismatch of the leading-order NRS scheme and the exact solution is improved by adding the residual viscous corrections. By adding terms of this kind one correctly matches the pressure anisotropy as seen in the right column of plots in Fig.~\ref{fig1}. The NLO NRS prescription agrees at high numerical precision with the predictions of the $\hat{\mathcal{P}}_L$ matching scheme. However, within the NLO NRS prescription one still needs the microscopic evolution of the anisotropy parameter $\xi$. On the other hand, for the $\hat{\mathcal{P}}_L$ matching scheme the equations of motion are formulated entirely in terms of the macroscopic quantities. 

\section{Conclusions}
In these proceedings we report the most recent developments of anisotropic hydrodynamics presented in Ref.~\cite{Martinez:2017ibh}. We compare the numerical solutions of the evolution equations corresponding to different hydrodynamical schemes for situations far from equilibrium with a recently found exact solution to the Boltzmann equation for the Gubser flow~\cite{Denicol:2014xca,Denicol:2014tha}. Our results show that anisotropic hydrodynamics with $\hat{\mathcal{P}}_L$ matching provides the most accurate description of the hydrodynamic moments obtained from the exact solution. Our results confirm similar findings obtained for the Bjorken flow~\cite{Molnar:2016gwq}. We also show that the leading order anisotropic hydrodynamics within the NRS prescription is improved by adding the residual dissipative term after which agrees almost perfectly with the $\hat{\mathcal{P}}_L$ matching scheme. One of the major advantages of the $\hat{\mathcal{P}}_L$ matching scheme is that, in contrast to all previous anisotropic hydrodynamical prescriptions, the total anisotropy pressure is matched exactly with the microscopic momentum anisotropy determined by the deformation parameter $\xi$. As a result, anisotropic hydrodynamics with $\hat{\mathcal{P}}_L$ matching can be formulated entirely in terms of macroscopic variables without any need to refer to the microscopic anisotropy parameter $\xi$. In this sense, the anisotropic hydrodynamical treatment is on the same footing as the traditional approaches to derive second order relativistic fluid dynamics. We conclude by mentioning that the $\hat{\mathcal{P}}_L$ matching scheme can be generalized to non-conformal systems undergoing an arbitrary expansion.

\vspace{3mm}

\noindent
{\bf Acknowledgments}:  
This work was supported by the U.S. Department of Energy, Office of Science, Office for Nuclear Physics under Awards DE-SC0004286 and DE-FG02-03ER41260 as well as within the framework of the Bean Energy Scan Theory (BEST) Topical Collaboration.
\vspace{-2mm}

\bibliographystyle{utphys}
\bibliography{QM17martinez}

\providecommand{\href}[2]{#2}\begingroup\raggedright\begin{thebibliography}{10}

\bibitem{Molnar:2016vvu}
E.~Molnar, H.~Niemi, and D.~H. Rischke, ``{Derivation of anisotropic
  dissipative fluid dynamics from the Boltzmann equation},''
  \href{http://dx.doi.org/10.1103/PhysRevD.93.114025}{{\em Phys. Rev.}
  {\bfseries D93} no.~11, (2016) 114025},
\href{http://arxiv.org/abs/1602.00573}{{ arXiv:1602.00573 [nucl-th]}}.

\bibitem{Molnar:2016gwq}
E.~Molnar, H.~Niemi, and D.~H. Rischke, ``{Closing the equations of motion of
  anisotropic fluid dynamics by a judicious choice of a moment of the Boltzmann
  equation},'' \href{http://dx.doi.org/10.1103/PhysRevD.94.125003}{{\em Phys.
  Rev.} {\bfseries D94} no.~12, (2016) 125003},
\href{http://arxiv.org/abs/1606.09019}{{ arXiv:1606.09019 [nucl-th]}}.

\bibitem{Jeon:2016uym}
S.~Jeon and U.~Heinz,
  \href{http://dx.doi.org/10.1142/9789814663717_0003}{``{Introduction to
  Hydrodynamics},''} in {\em Quark-Gluon Plasma 5}, X.-N. Wang, ed.,
  pp.~131--187.
\newblock
2016.
\newblock

\bibitem{Martinez:2010sc}
M.~Martinez and M.~Strickland, ``{Dissipative Dynamics of Highly Anisotropic
  Systems},'' \href{http://dx.doi.org/10.1016/j.nuclphysa.2010.08.011}{{\em
  Nucl. Phys.} {\bfseries A848} (2010) 183--197},
\href{http://arxiv.org/abs/1007.0889}{{ arXiv:1007.0889 [nucl-th]}}.

\bibitem{Martinez:2010sd}
M.~Martinez and M.~Strickland, ``{Non-boost-invariant anisotropic dynamics},''
  \href{http://dx.doi.org/10.1016/j.nuclphysa.2011.02.003}{{\em Nucl. Phys.}
  {\bfseries A856} (2011) 68--87},
\href{http://arxiv.org/abs/1011.3056}{{ arXiv:1011.3056 [nucl-th]}}.

\bibitem{Florkowski:2010cf}
W.~Florkowski and R.~Ryblewski, ``{Highly-anisotropic and strongly-dissipative
  hydrodynamics for early stages of relativistic heavy-ion collisions},''
  \href{http://dx.doi.org/10.1103/PhysRevC.83.034907}{{\em Phys. Rev.}
  {\bfseries C83} (2011) 034907},
\href{http://arxiv.org/abs/1007.0130}{{ arXiv:1007.0130 [nucl-th]}}.

\bibitem{Ryblewski:2010bs}
R.~Ryblewski and W.~Florkowski, ``{Non-boost-invariant motion of dissipative
  and highly anisotropic fluid},''
  \href{http://dx.doi.org/10.1088/0954-3899/38/1/015104}{{\em J. Phys.}
  {\bfseries G38} (2011) 015104},
\href{http://arxiv.org/abs/1007.4662}{{ arXiv:1007.4662 [nucl-th]}}.

\bibitem{Bazow:2013ifa}
D.~Bazow, U.~Heinz, and M.~Strickland, ``{Second-order (2+1)-dimensional
  anisotropic hydrodynamics},''
  \href{http://dx.doi.org/10.1103/PhysRevC.90.054910}{{\em Phys. Rev.}
  {\bfseries C90} no.~5, (2014) 054910},
\href{http://arxiv.org/abs/1311.6720}{{ arXiv:1311.6720 [nucl-th]}}.

\bibitem{Denicol:2014loa}
G.~S. Denicol, ``{Kinetic foundations of relativistic dissipative fluid
  dynamics},''
\href{http://dx.doi.org/10.1088/0954-3899/41/12/124004}{{\em J. Phys.}
  {\bfseries G41} no.~12, (2014) 124004}.

\bibitem{Romatschke:2003ms}
P.~Romatschke and M.~Strickland, ``Collective modes of an anisotropic quark
  gluon plasma,'' {\em Phys. Rev.} {\bfseries D68} (2003) 036004,
\href{http://arxiv.org/abs/hep-ph/0304092}{{ hep-ph/0304092}}.

\bibitem{Martinez:2017ibh}
M.~Martinez, M.~McNelis, and U.~Heinz, ``{Anisotropic fluid dynamics for Gubser
  flow},''
\href{http://arxiv.org/abs/1703.10955}{{ arXiv:1703.10955 [nucl-th]}}.

\bibitem{Gubser:2010ze}
S.~S. Gubser, ``{Symmetry constraints on generalizations of Bjorken flow},''
  \href{http://dx.doi.org/10.1103/PhysRevD.82.085027}{{\em Phys. Rev.}
  {\bfseries D82} (2010) 085027},
\href{http://arxiv.org/abs/1006.0006}{{ arXiv:1006.0006 [hep-th]}}.

\bibitem{Gubser:2010ui}
S.~S. Gubser and A.~Yarom, ``{Conformal hydrodynamics in Minkowski and de
  Sitter spacetimes},''
  \href{http://dx.doi.org/10.1016/j.nuclphysb.2011.01.012}{{\em Nucl. Phys.}
  {\bfseries B846} (2011) 469--511},
\href{http://arxiv.org/abs/1012.1314}{{ arXiv:1012.1314 [hep-th]}}.

\bibitem{Denicol:2014xca}
G.~S. Denicol, U.~Heinz, M.~Martinez, J.~Noronha, and M.~Strickland, ``{New
  Exact Solution of the Relativistic Boltzmann Equation and its Hydrodynamic
  Limit},'' \href{http://dx.doi.org/10.1103/PhysRevLett.113.202301}{{\em Phys.
  Rev. Lett.} {\bfseries 113} no.~20, (2014) 202301},
\href{http://arxiv.org/abs/1408.5646}{{ arXiv:1408.5646 [hep-ph]}}.

\bibitem{Denicol:2014tha}
G.~S. Denicol, U.~Heinz, M.~Martinez, J.~Noronha, and M.~Strickland,
  ``{Studying the validity of relativistic hydrodynamics with a new exact
  solution of the Boltzmann equation},''
  \href{http://dx.doi.org/10.1103/PhysRevD.90.125026}{{\em Phys. Rev. D}
  {\bfseries 90} no.~12, (2014) 125026},
\href{http://arxiv.org/abs/1408.7048}{{ arXiv:1408.7048 [hep-ph]}}.

\bibitem{Denicol:2012cn}
G.~S. Denicol, H.~Niemi, E.~Molnar, and D.~H. Rischke, ``{Derivation of
  transient relativistic fluid dynamics from the Boltzmann equation},''
  \href{http://dx.doi.org/10.1103/PhysRevD.85.114047,
  10.1103/PhysRevD.91.039902}{{\em Phys. Rev.} {\bfseries D85} (2012) 114047},
  \href{http://arxiv.org/abs/1202.4551}{{ arXiv:1202.4551 [nucl-th]}}.
[Erratum: Phys. Rev.D91,no.3,039902(2015)].

\bibitem{Nopoush:2014qba}
M.~Nopoush, R.~Ryblewski, and M.~Strickland, ``{Anisotropic hydrodynamics for
  conformal Gubser flow},''
  \href{http://dx.doi.org/10.1103/PhysRevD.91.045007}{{\em Phys. Rev. D}
  {\bfseries 91} no.~4, (2015) 045007},
\href{http://arxiv.org/abs/1410.6790}{{ arXiv:1410.6790 [nucl-th]}}.

\end{thebibliography}\endgroup

\end{document}